\documentclass[aps,amsmath]{revtex4}
\usepackage{graphicx} 

\renewcommand{\Im}{\mbox{Im}}

\renewcommand{\i}{{\rm i}}
\renewcommand{\d}{{\rm d}}

        \newcommand{\la}{\langle}
        \newcommand{\ra}{\rangle}

        \newcommand{\qbar}{{\bar q}}
        \newcommand{\bse}{\begin{subequations}}
        \newcommand{\ese}{\end{subequations}}
        
        \newcommand{\ben}{\begin{eqnarray*}}
        \newcommand{\een}{\end{eqnarray*}}
        \newcommand{\be}{\begin{equation}}
        \newcommand{\ee}{\end{equation}}
        \newcommand{\half}{\frac{1}{2}}
        \newcommand{\tr}{\mbox{Tr}}

        \newcommand{\lek}{\left(k\right)}
        \newcommand{\lel}{\left(l\right)}

\begin{document}

\preprint{}

\title{Vector and axialvector mesons at nonzero temperature 
within a gauged linear sigma model}

\author{Stefan Str\"uber}
\email{strueber@th.physik.uni-frankfurt.de}
\affiliation{Institut f\"ur Theoretische Physik,
Johann Wolfgang Goethe-Universit\"at, \\
Max von Laue-Str.\ 1, D-60438 Frankfurt/Main, Germany}
\author{Dirk H.\ Rischke}
\email{drischke@th.physik.uni-frankfurt.de}
\affiliation{Institut f\"ur Theoretische Physik and
Frankfurt Institute for Advanced Studies,\\
Johann Wolfgang Goethe-Universit\"at, 
Max von Laue-Str.\ 1, D-60438 Frankfurt/Main, Germany}
\begin{abstract}
We consider vector and axialvector mesons in the framework of
a gauged linear sigma model with chiral
$U(N_f)_R \times U(N_f)_L$ symmetry. For $N_f=2$, we investigate
the behavior of the chiral condensate and the meson masses 
as a function of temperature by solving a system of coupled 
Dyson-Schwinger equations derived
via the 2PI formalism in double-bubble
approximation. We find that the
inclusion of vector and axialvector mesons tends to sharpen the
chiral transition. Within our approximation scheme, the 
mass of the $\rho$ meson increases by about 100 MeV 
towards the chiral transition.
\end{abstract}

\date{\today}
\maketitle


\section{Introduction}\label{I}

The fundamental theory of the strong interaction is
quantum chromodynamics (QCD). QCD has
a local $SU(3)_c$ gauge symmetry which determines the
interaction between matter constituents, the quarks, and 
gauge fields, the gluons. For massless quarks,
the quark sector of the QCD Lagrangian also has a
global $U(N_f)_R \times U(N_f)_L$ chiral symmetry, where
$N_f$ is the number of quark flavors.
The $U(1)_A$ anomaly induced by instantons 
\cite{'tHooft:1986nc}
breaks this symmetry explicitly to 
$U(N_f)_V \times SU(N_f)_A  \times Z(N_f)_A$, where
vector and axial vector symmetries are introduced
via $V = R+L, \, A=R-L$. 
The discrete $Z(N_f)_A$ symmetry plays no role
for the dynamics and will be omitted in the following.
Nonzero, degenerate quark masses explicitly break 
$SU(N_f)_A$, such that the remaining
symmetry is $U(N_f)_V$.
Non-degenerate quark masses further break this
symmetry to $U(1)_V$, 
corresponding to baryon number conservation.

At low momenta $\sim \Lambda_{\rm QCD}$, where
$\Lambda_{\rm QCD} \sim 200$ MeV is the QCD scale parameter,
quarks and gluons are confined
inside hadrons. Therefore, on a typical hadronic
length scale $\sim \Lambda_{\rm QCD}^{-1} \sim 1$ fm,
the $SU(3)_c$ gauge symmetry of QCD is (at best) of minor 
importance, and the interactions between hadrons are
predominantly 
determined by the global $U(N_f)_R \times U(N_f)_L$
chiral symmetry of QCD.
In the QCD vacuum, the axial $U(N_f)_A$ part of the
latter symmetry is spontaneously
broken by a non-vanishing expectation value of the 
quark condensate $\la \qbar q \ra \neq 0$
\cite{Vafa:1983tf}. According to Goldstone's theorem,
this would lead to $N_f^2$ Goldstone bosons. However,
since the explicit symmetry breaking induced by the
$U(1)_A$ anomaly reduces the axial symmetry to
$SU(N_f)_A$, spontaneous symmetry breaking
of the latter gives rise to only $N_f^2-1$ Goldstone bosons.
These Goldstone bosons acquire a mass due to the explicit
chiral symmetry breaking by nonzero quark masses.

At temperatures of the order of 
$\sim\la\qbar q\ra^{1/3}$, the thermal 
excitation energy is large enough to restore the 
$SU(N_f)_A$ symmetry of QCD. If instantons
are sufficiently screened \cite{Gross:1980br}, 
this could additionally lead to
a restoration of the explicitly broken $U(1)_A$.
For vanishing quark masses, the high- and the low-temperature
phases of QCD have different symmetries, and therefore must
be separated by a phase transition. 
The order of this chiral phase
transition is determined by the global symmetry of
the QCD Lagrangian;
for $U(N_f)_R \times U(N_f)_L$, the transition
is of first order if $N_f \geq 2$, for
$SU(N_f)_R \times SU(N_f)_L \times U(1)_V$, the
transition can be of second order if $N_f \leq 2$
\cite{Pisarski:1983ms}.
If the quark masses are nonzero, the second-order 
phase transition becomes cross-over.

Lattice QCD calculations predict the chiral
phase transition to happen at a
temperature $T_c \sim 150 - 190$ MeV
\cite{Aoki:2006br,Karsch:2007vw} 
for zero quark chemical potential $\mu$. The phase transition
temperature is expected to decrease when $\mu$ increases
and vanishes at some value $\mu_c$ corresponding to
quark number densities of the order of a few times
nuclear matter density.
Therefore, the chiral transition of QCD is the only
phase transition in a theory of one of the fundamental
forces of nature which can be studied under
laboratory conditions: 
heavy-ion collision experiments performed at the accelerator
facilities CERN-SPS,
BNL-RHIC and, in the near future, CERN-LHC and
GSI-FAIR create matter which is sufficiently hot and/or dense, 
such that chiral symmetry is restored. Indeed, the primary
goal of these experiments is to find 
evidence for the restoration of chiral symmetry by
the creation of the so-called quark-gluon plasma, i.e., the
phase of QCD where quarks and gluons are liberated
from confinement.

When chiral symmetry is restored, the masses of
hadrons with the same quantum numbers except for
parity and G-parity, so-called chiral partners, 
become degenerate. Chiral partners are, for instance,
the sigma and the pion in the (pseudo-)scalar sector, or 
the $\rho$ and the $a_1$ in the (axial) vector sector. 
A promising signal for chiral
symmetry restoration in heavy-ion collisions is
therefore to study changes of the spectral properties
of hadrons in the hot and dense environment 
\cite{Pisarski:1994yp,Brown:2005kb,Rapp:1999ej}. One of the prime
candidates is the $\rho$ meson. The $\rho$ meson decays
sufficiently fast (and with -- for experimental purposes --
sufficiently large branching ratio) 
into a pair of dileptons which, due to
their small (since electromagnetic) cross section, 
are able to carry information from the hot and 
dense stages of a heavy-ion collision to the detector.
The CERES and NA60 experiments at the CERN-SPS have
found convincing evidence for a modification
of the $\rho$ meson spectral function in
Pb+Pb and In+In collisions, respectively 
\cite{Agakichiev:2005ai,Damjanovic:2007qm}.

It is important to clarify whether the modification of the
$\rho$ meson spectral function observed by the
CERN-SPS experiments is in any way related to
chiral symmetry restoration or is merely due
to many-body interactions in the hot and dense medium.
This question can be decided by calculating the dilepton
production rate from QCD and
then using this rate in a dynamical model for 
heavy-ion collisions in order to compute the
dilepton spectrum. The low-invariant mass region
of the dilepton spectrum is dominated by the decay of
hadronic states. Therefore, for the calculation of the
dilepton rate it is more convenient to apply a 
low-energy effective theory for QCD, featuring hadronic
states as elementary degrees of freedom and respecting
the chiral symmetries of QCD, rather than using QCD itself.
Since we are interested in the restoration of
chiral symmetry at nonzero temperature, the low-energy
effective theory of choice is a linear sigma model
which treats hadrons and their chiral partners 
on the same footing.

Linear sigma models have been used for quite
some time in order to study chiral
symmetry restoration in hot and dense strongly
interacting matter. For instance, Pisarski and Wilczek have
applied renormalization group arguments to
a $U(N_f)_V\times U(N_f)_A$ symmetric linear sigma model 
with scalar degrees of freedom and have drawn important
qualitative conclusions regarding the order of
the chiral phase transition for different numbers
of quark flavors \cite{Pisarski:1983ms}.
The calculation of hadronic properties at 
nonzero temperature (and density) faces serious
technical difficulties. For instance, for the following
reasons it is impossible to apply standard perturbation
theory. First, it turns out that the coupling constants
of effective low-energy models of QCD are of the order of one, 
rendering a perturbative series in the coupling constant
unreliable. Second, nonzero temperature (or density)
introduces an additional scale which invalidates
the usual power counting in terms of the coupling constant
\cite{Dolan:1973qd}. A consistent calculation to a
given order in the coupling constant then may require 
a resummation of whole classes of diagrams \cite{Braaten:1989mz}.

A convenient technique to perform such a
resummation and thus arrive at a particular 
many-body approximation scheme is the so-called
two-particle irreducible (2PI) or Cornwall-Jackiw-Tomboulis 
(CJT) formalism \cite{Cornwall:1974vz}, which is a relativistic 
generalisation of the $\Phi$ functional 
formalism \cite{Luttinger:1960ua,Baym:1962sx}. 
The 2PI formalism extends the concept of 
the generating functional $\Gamma[\phi]$ for one-particle
irreducible (1PI) Green's functions to that of one for
2PI Green's functions
$\Gamma[\phi,G]$, where $\phi$ and $G$ are the 
expectation values of the one- and two-point functions. 
The central quantity in this formalism is the sum of 
all 2PI vacuum diagrams, $\Gamma_2[\phi,G]$. Any
many-body approximation scheme can be derived as a 
particular truncation of $\Gamma_2[\phi,G]$. 

An advantage of the 2PI formalism is that it avoids 
double counting and fulfills detailed balance relations and 
thus is thermodynamically consistent. 
Another advantage is that the Noether currents are 
conserved for an arbitrary truncation of 
$\Gamma_2$ as long as the one- and 
two-point functions transform
as rank-1 and -2 tensors. A disadvantage 
is that Ward-Takahashi identities for higher-order 
vertex functions are no longer fulfilled \cite{vanHees:2002bv}. 
As a consequence, Goldstone's theorem is violated
\cite{Petropoulos:1998gt,Lenaghan:1999si}. 
Another consequence is that sum rules of the Weinberg type 
at zero and nonzero temperature \cite{Kapusta:1993hq}
are not necessarily fulfilled.
A strategy to 
restore Goldstone's theorem is to perform a so-called
``external'' resummation of random-phase-approximation 
diagrams with internal lines given by the full propagators 
of the underlying approximation in the 2PI formalism
\cite{vanHees:2002bv}. In this work, however, this problem
is less severe since we shall only focus on the case
of explicit chiral symmetry breaking by (small) non-vanishing
quark masses.

Effective theories with chiral
$U(N_f)_R\times U(N_f)_L$ and $O(N)$ symmetry have already been 
studied within the 2PI formalism in the so-called
double-bubble approximation which gives rise to
Hartree-Fock-type self-energies for the individual particles
\cite{Baym:1977qb,Bochkarev:1995gi,Polchinski:1996na,Amelino-Camelia:1997dd,Petropoulos:1998gt,Lenaghan:1999si,Lenaghan:2000ey,Roder:2003uz}. The $O(N)$ model has also been studied
within the 2PI formalism 
beyond the double-bubble approximation 
including sunset-type diagrams \cite{Roder:2005vt}.
The chiral models in these works, however, only include 
scalar and pseudoscalar particles. Here, we extend these 
investigations to vector and axial vector mesons. 

Our ultimate goal is a calculation
of the dilepton rate for given temperature (and density).
In this work, we perform the first step in this direction
studying chiral symmetry restoration in the mesonic mass 
spectrum at nonzero temperature.
We apply the so-called gauged linear sigma model as outlined in
Ref.\ \cite{Gasiorowicz:1969kn}.
This model has been used by Pisarski \cite{Pisarski:1994yp}
who varied the tree-level mass parameters in order to derive
qualitative statements about the behavior of meson
masses as a function of temperature. Here, we extend
these studies 
to a full-fledged self-consistent many-body
calculation of meson masses at nonzero temperature using
the 2PI formalism in double-bubble approximation. 

This paper is organized as follows.
The following section is dedicated to a discussion
of the gauged linear sigma model with
$U(N_f)_R \times U(N_f)_L$ chiral symmetry, with
$N_f$ being the number of quark flavors.
In Sec.\ \ref{III} we consider the
case $N_f=2$ in more detail and rewrite the Lagrangian
explicitly in terms of scalar, pseudoscalar, vector,
and axial vector degrees of freedom. We also discuss
issues related to the breaking of chiral symmetry such
as the mixing of pseudoscalar and axial vector mesons.
In Sec.\ \ref{IV} we use this model to 
calculate the behavior of the meson masses and the 
condensate as a function of temperature.
Section \ref{V} concludes this paper with a summary 
of our results and an outlook for further studies.

We use the imaginary-time formalism to compute quantities at 
nonzero temperature. Our notation is
\begin{equation}
\int_k f(k)=T \sum_{n=-\infty}^{\infty} 
 \int  \frac{\d^3 k}{\left(2\pi\right)^3}f\left(2\pi\i nT,
  \vec k\right)\,\,.
\end{equation}
We use units $\hbar=c=k_B=1$. The metric tensor is 
$g^{\mu\nu}=\mbox{diag}(1,-1,-1,-1)$.

\section{The gauged linear sigma model}\label{II}

In this section we present the gauged linear sigma model
with $U(N_f)_R \times U(N_f)_L$ symmetry.
The Lagrangian reads \cite{Gasiorowicz:1969kn}
\begin{eqnarray}
{\cal L} & = & 
\mbox{Tr}[(D^{\mu} \Phi )^{\dag}D_{\mu} \Phi ]-m^{2}_{0}
\mbox{Tr}[\Phi^{\dag}\Phi]
-\lambda_{1}(\mbox{Tr}[\Phi^{\dag}\Phi])^{2}-
\lambda_{2}\mbox{Tr}[(\Phi^{\dag}\Phi)^{2}]
-\frac{1}{4} \mbox{Tr}[(L^{\mu\nu})^{2}+(R^{\mu\nu})^{2}]
\nonumber\\
&&+\frac{m^{2}_{1}}{2}\mbox{Tr}[(L^{\mu})^{2}+(R^{\mu})^{2}]
+c \left[ \mbox{det}(\Phi)+\mbox{det}(\Phi^{\dag}) \right]
+\mbox{Tr}[H(\Phi+\Phi^{\dag})]
\label{vecscal}\;.
\end{eqnarray}
Here, the field $\Phi$ stands for scalar and pseudoscalar
hadronic degrees. These are chosen to be organized into 
representations $[N_f^*,N_f]$ under
$U(N_f)_R \times U(N_f)_L$ transformations, such that
\begin{equation}
\Phi \rightarrow  U_R^\dag \, \Phi U_L\;,
\end{equation}
where $U_{R,L}$ are unitary matrices acting on the fundamental
representation of $U(N_f)_{R,L}$.
This allows for an explicit
representation of $\Phi$ in terms of a complex 
$N_f \times N_f$ matrix,
\begin{equation}
\Phi=\left(\sigma_a+\i\pi_a\right)T_a\,\mbox{,}
\end{equation}
where $T_a,\, a=0, \ldots, N_f^2-1$, are the generators
of $U(N_f)$ in the fundamental representation, and
$\sigma_a$, $\pi_a$, parametrize
scalar and pseudoscalar fields, respectively. 

In order to introduce vector and axialvector 
degrees of freedom, one defines right- and
left-handed vector fields, 
\begin{equation}
R^\mu = (V^\mu_a + A^\mu_a) T_a\;, \;\;\;\;
L^\mu = (V^\mu_a - A^\mu_a) T_a\;,
\end{equation}
where $V^\mu_a$, $A^\mu_a$ are vector and axialvector fields,
respectively.
These are treated as massive Yang-Mills fields, with 
field-strength tensor
\begin{equation}
R^{\mu \nu} = \partial^{\mu} R^{\nu}- 
\partial^{\nu} R^{\mu}-{\rm i}g\left[R^{\mu},R^{\nu}\right] 
\; , \;\;\;
L^{\mu \nu} = \partial^{\mu} L^{\nu}- 
\partial^{\nu} L^{\mu}-{\rm i}g\left[L^{\mu},L^{\nu}\right]\;.
\end{equation}
Right- and left-handed fields transform as gauge fields under
$U(N_f)_{R,L}$,
\begin{equation}
R^\mu \rightarrow U_R^\dag \left(R^\mu + \frac{{\rm i}}{g}\,
\partial^\mu \right) U_R\;, \;\;\;\;
L^\mu \rightarrow U_L^\dag \left(L^\mu + \frac{{\rm i}}{g}\, 
\partial^\mu \right) U_L\;, 
\end{equation}
while the field-strength tensors transform covariantly under
$U(N_f)_{R,L}$,
\begin{equation}
R^{\mu \nu} \rightarrow U_R^\dag R^{\mu \nu} U_R\;, \;\;\;\;
L^{\mu \nu} \rightarrow U_L^\dag L^{\mu \nu} U_L\;.
\end{equation}
The covariant derivative in Eq.\ (\ref{vecscal}) couples 
scalar and pseudoscalar degrees
of freedom to right- and left-handed vector fields,
\begin{equation}
D_\mu \Phi = \partial_\mu \Phi + {\rm i} g (\Phi L_\mu - 
R_\mu \Phi)\;.
\end{equation}

The first line in Eq.\ (\ref{vecscal}) contains terms which are
invariant under {\em local\/} $U(N_f)_R \times U(N_f)_L$
transformations. The vector meson mass term renders this 
symmetry a global one. The determinant terms represent 
the $U(1)_A$ anomaly of QCD and break the symmetry 
explicitly to $SU(N_f)_R \times SU(N_f)_L \times U(1)_V$, 
where $U(1)_V$ stands for baryon number conservation. 
The last term in Eq.\ (\ref{vecscal}) corresponds
to the quark mass term in the QCD Lagrangian. Since
this term is flavor-diagonal, we have
$H= \sum_{i=1}^{N_f} h_{i^2-1} T_{i^2-1}$. For degenerate 
nonzero quark masses, $h_0 \neq 0$, while all other 
$h_{i^2-1}$ vanish. In this case, $U(N_f)_R \times U(N_f)_L$ 
is explicitly broken to $U(N_f)_V$. For non-degenerate 
quark masses, also the other $h_{i^2-1}$ have to be nonzero. 
Exact $SU(2)_V$ isospin symmetry requires $h_3 = 0$.
 
Local $U(N_f)_R \times U(N_f)_L$ invariance implies
universality of the coupling, i.e., the coupling $g$ between 
right- and left-handed vector fields to scalar and 
pseudoscalar fields is the same as the coupling of the 
vector fields among themselves. Note that the Lagrangian
(\ref{vecscal}) would also be locally invariant under 
$U(N_f)_R \times U(N_f)_L$, had we introduced 
{\em separate\/} coupling constants $g_R,\, g_L$ for right- and
left-handed vector fields and modified the covariant derivative,
$D_{\mu}\Phi=\partial_{\mu}\Phi+{\rm i}(g_L \Phi L_{\mu}- 
g_R R_{\mu}\Phi)$. However, if $g_R \neq g_L$, one 
obtains non-vanishing parity-violating terms $\sim g_R - g_L$ 
in the Lagrangian, which must not occur in viable 
theories of the strong interaction.

The assumption of local invariance further restricts 
possible couplings between scalar and vector mesons. Under
global transformations, terms like $\Phi L^{\mu}$ and 
$R^{\mu} \Phi$ would transform as $\Phi$ itself.
Taking discrete symmetries into
account possible coupling terms would be 
${\rm Tr}\, |\Phi L^{\mu}-R^{\mu} \Phi|^2$, 
${\rm Tr}\, |\Phi L^{\mu}+R^{\mu}\Phi|^2$ and 
${\rm Tr}\,\left(|\Phi L^{\mu}|^2+|R^{\mu}\Phi|^2\right)$
\cite{Pisarski:1994yp}. 

According to Noether's theorem, continuous symmetries lead to
conserved currents. Explicit symmetry breaking induces
source terms in the conservation laws.
For global symmetries, there is a simple way to derive the 
currents and the conservation laws
\cite{Gell:1960}. Consider the symmetry transformation 
to be of the form $U=\exp(- {\rm i} \theta_a T_a)$.
Promoting the global symmetry to a local one, 
$\theta_a \rightarrow \theta_a(X)$, and computing
the variation $\delta {\cal L}$ of the Lagrangian,
one can then read off the Noether currents ${\cal J}^\mu_a$ 
and the conservation laws from the identity
\begin{equation}
\delta {\cal L} = - \partial_\mu {\cal J}^\mu_a \theta_a
- {\cal J}^\mu_a\, \partial_\mu \theta_a\;.
\end{equation}
For QCD, the $U(N_f)_R \times U(N_f)_L = 
U(N_f)_V \times U(N_f)_A$ symmetry leads to the
following vector and axial-vector currents and the 
conservation laws:
\begin{subequations}
\begin{eqnarray}
{\cal{J}}_{V\mu}^a&=&- \frac{\partial\delta{\cal L}_{QCD}}{
                           \partial(\partial_{\mu}\theta_V^a)}
         =\bar q\gamma^{\mu}T_aq\,\,\mbox{,}\\
{\cal{J}}_{A\mu}^a&=&-\frac{\partial\delta{\cal L}_{QCD}}{
                           \partial(\partial_{\mu}\theta_A^a)}
         =\bar q\gamma^{\mu}\gamma_5T_aq\,\mbox{,} \\
\partial_{\mu}{\cal{J}}^{a\mu}_V&=&
  - \frac{\partial\delta{\cal L}_{QCD}}{\partial\theta_V^a}
         ={\rm i}\bar q\left[M,T_a\right]q\,\,\mbox{,}\\
\partial_{\mu}{\cal{J}}^{a\mu}_A &=& 
   - \frac{\partial\delta{\cal L}_{QCD}}{\partial\theta_A^a}
         ={\rm i}\bar q\left\{T_a,M\right\}\gamma_5 q
          +\delta_{a0}\, \frac{g^2N_f}{64\pi^2} 
          \epsilon_{\mu \nu \alpha \beta}\,
{G}_b^{\mu\nu}G_b^{\alpha\beta} \,\mbox{,}
\end{eqnarray}
\end{subequations}
where $M$ is the quark mass matrix and $G_b^{\mu \nu}$ the gluon
field-strength tensor. The last term on the right-hand
side of the conservation law for the axial current cannot 
be obtained from the variation of the classical QCD Lagrangian. 
It represents the $U(1)_A$ anomaly and arises from 
instantons \cite{'tHooft:1986nc}.

In the gauged linear sigma model, the vector and 
axial-vector currents and the corresponding conservation 
laws can be obtained analogously from the variation of the 
Lagrangian (\ref{vecscal}) under local 
$U(N_f)_V \times U(N_f)_A$ transformations. The result is
\begin{subequations}
\begin{eqnarray}
{\cal{J}}_{V\mu}^a&=& -
\frac{\partial\delta{\cal L}}{\partial(\partial_{\mu}\theta_V^a)}
        = \frac{m^2_1}{g}V^a_{\mu}\,\,\mbox{,}\qquad\,\,\,\;
{\cal{J}}_{A\mu}^a= -
\frac{\partial\delta{\cal L}}{\partial(\partial_{\mu}\theta_A^a)}
        = \frac{m^2_1}{g}A^a_{\mu}\,\mbox{,}
\label{noeth}\\
\partial_{\mu}{\cal{J}}^{a\mu}_V&=& -
         \frac{\partial\delta{\cal L}}{\partial\theta_V^a}
        = - f_{abc}\sigma_bh_c\,\,\mbox{,}\qquad\,\,\,\,
\partial_{\mu}{\cal{J}}^{a\mu}_A = -
         \frac{\partial\delta{\cal L}}{\partial\theta_A^a}
        = d_{abc}\pi_bh_c
    + 4 \delta_{a0}\, c\, 
     \Im\left[\mbox{det}\left(\Phi\right)\right] \,\,\mbox{,}
\label{conserve}
\end{eqnarray}
\end{subequations}
where $f_{abc}, d_{abc}$ are
the totally antisymmetric and
symmetric structure constants of $SU(N)$, respectively.
Since the first line in Eq.\ (\ref{vecscal}) is {\em locally\/}
invariant under $U(N_f)_V \times U(N_f)_A$, it cannot 
contribute to the currents or the conservation laws. 
Then, the vector meson mass term which violates local 
$U(N_f)_V \times U(N_f)_A$ invariance gives rise to the
celebrated current-field proportionality (\ref{noeth}) 
\cite{Gasiorowicz:1969kn}.

One may wonder why the vector current, 
$\tilde{\cal J}_{V\mu}^a$, 
and axial vector current, $\tilde{\cal J}_{A\mu}^a$, 
arising from scalar and pseudoscalar particles
do not appear in the expressions (\ref{noeth}) 
for the vector and axial vector currents. They read
explicitly
\begin{subequations}
\begin{eqnarray}
\tilde{\cal J}_{V \mu}^a & = & - \frac{1}{2}\, f_{abc}
\left[ \phi_b^* (D_\mu \phi)_c - (D_\mu \phi)_b^* \phi_c
\right]\;, \\
\tilde{\cal J}_{A \mu}^a & = & - \frac{{\rm i}}{2}\, d_{abc}
\left[ \phi_b^* (D_\mu \phi)_c - (D_\mu \phi)_b^* \phi_c
\right]\;,
\end{eqnarray}
\end{subequations} 
where
\begin{equation}
(D_\mu \phi)_a  \equiv  \left[ \partial_\mu \delta_{ac}
+g \left( f_{abc} V_{\mu}^b - {\rm i} d_{abc} A_\mu^b
\right) \right] \phi_c
\end{equation}
is the covariant derivative.
However, employing the equations of motion,
these currents are recognized as simply being parts of the total
vector and axial vector currents (\ref{noeth}),
\begin{subequations}
\begin{eqnarray}
{\cal J}_{V\mu}^a
& = & \tilde{\cal J}_{V \mu}^a - \frac{1}{g}
\, \partial^\nu V_{\nu \mu}^a -  f_{abc} \left(
V_b^\nu V_{\nu \mu}^c + A^{\nu}_b A_{\nu \mu}^c \right) 
\;,\\
{\cal J}_{A\mu}^a
& = & \tilde{\cal J}_{A \mu}^a - \frac{1}{g}\,
\partial^\nu A_{\nu \mu}^a -  f_{abc} \left(
V_b^\nu A_{\nu \mu}^c + A^{\nu}_b V_{\nu \mu}^c \right) 
\;,
\end{eqnarray}
\end{subequations}
where
\begin{subequations}
\begin{eqnarray}
V_{\mu \nu}^a & \equiv & \partial_\mu V_\nu^a - \partial_\nu
V_\mu^a + g f_{abc} \left( V_\mu^b V_\nu^c + A_\mu^b A_\nu^c
\right) \;, \\
A_{\mu \nu}^a & \equiv & \partial_\mu A_\nu^a - \partial_\nu
A_\mu^a + g f_{abc} \left( V_\mu^b A_\nu^c + A_\mu^b V_\nu^c
\right)
\end{eqnarray}
\end{subequations}
are the field strength tensors for vector and axial vector
fields, respectively.

In the following, we briefly discuss consequences of Eqs.\
(\ref{conserve}). In the chiral limit, $h_a =0$, and
the isosinglet vector current ${\cal J}_{V\mu}^0$ and the
vector currents ${\cal J}_{V\mu}^i, \, i\neq 0$,
as well as the axial vector currents ${\cal J}_{A\mu}^i$, 
are exactly conserved. The isosinglet axial vector current 
${\cal J}_{A\mu}^0$ receives a contribution from the 
$U(1)_A$ anomaly. In the case of explicit chiral symmetry 
breaking, we have $h_a\neq 0$. However, since  $f_{0bc}=0$,
the isosinglet vector current is still exactly conserved.
If only $h_0 \neq 0$, but $h_i=0,\, i \neq 0$,
for the same reason also the other vector currents 
${\cal J}_V^i$ are exactly conserved.
For instance, this is the case for $N_F=2$ assuming isospin
invariance which requires $h_i=0$ for $i=1,2,3$.
In contrast, since $d_{ab0} = \delta_{ab}$, the axial vector
currents are only partially conserved; 
the famous partial conservation of axial currents (PCAC).

\section{The case {\boldmath $N_f=2$}}\label{III}

So far, the discussion of the Lagrangian (\ref{vecscal}) 
was valid for an arbitrary number of quark flavours. 
In the following, we restrict ourselves to the case of 
mass-degenerate up and down quarks. For this case, 
the fields $\Phi$, $R^\mu$, and $L^\mu$ can be written in 
terms of the physical scalar ($\sigma,\,\vec{a}_0$), 
pseudoscalar ($\eta, \, \vec{\pi}$), as well as vector
($\omega^\mu,\, \vec{\rho}^{\,\mu}$) and axial vector 
fields ($f_1^\mu,\, \vec{a}_1^{\,\mu}$) as:
\begin{subequations}
\begin{eqnarray}
\Phi&=&\left(\sigma_a+\i\pi_a\right)t_a=
       \left(\sigma+\i\eta\right)t_0
       +\left(\vec{a}_0+\i\vec{\pi}\right)\cdot \vec{t}\; ,\\
R^{\mu}&=&\left(V^{\mu}_a+A^{\mu}_a\right)t_a
        =\left(\omega^{\mu}+f_1^{\mu}\right)t_0
   +\left(\vec{\rho}^{\, \mu}
   +\vec{a}_1^{\, \mu}\right)\cdot \vec{t}
\; ,\\
L^{\mu}&=&\left(V^{\mu}_a-A^{\mu}_a\right)t_a
        =\left(\omega^{\mu}-f_1^{\mu}\right)t_0
   +\left(\vec{\rho}^{\, \mu}
   -\vec{a}_1^{\, \mu}\right)\cdot \vec{t}
\,\,\mbox{.}
\end{eqnarray}
\end{subequations}
The vector and axial vector mesons enter the Lagrangian 
through the following terms:
\begin{subequations}
\begin{eqnarray}
\mbox{Tr}[(D^{\mu} \Phi )^{\dag}D_{\mu} \Phi ]&=&
 \;\;\;\frac{1}{2} \left[\partial_{\mu}\sigma+
  g(\eta\, f_{1\mu} + \vec{\pi}\cdot\vec{a}_{1\mu}) \right]^{2}
\nonumber\\
 & & +\frac{1}{2} \left[\partial_{\mu}\eta-
  g(\sigma\,f_{1\mu}+\vec{a}_{0}\cdot \vec{a}_{1\mu})\right]^{2} 
\nonumber \\ 
 & & +\frac{1}{2} \left[\partial_{\mu}\vec{a}_{0}+
  g(\vec{\rho}_{\mu}\times \vec{a}_{0} +\eta\, \vec{a}_{1\mu} 
   +\vec{\pi}\,f_{1\mu})\right]^{2} 
\nonumber \\
 & & +\frac{1}{2}\left[\partial_{\mu}\vec{\pi}-g(\vec{\pi} \times
      \vec{\rho}_{\mu} +\sigma\, \vec{a}_{1\mu}
     +\vec{a}_{0}\,f_{1\mu})\right]^{2} \,\mbox{,} \label{vs}\\
 - \frac{1}{4} \mbox{Tr}[(L^{\mu\nu})^{2}+(R^{\mu\nu})^{2}]&=&-
  \frac{1}{4}(\partial_{\mu}\omega_{\nu}
  -\partial_{\nu}\omega_{\mu})^{2}  
\nonumber\\
 & & -\frac{1}{4}(\partial_{\mu}f_{1\nu}
                 -\partial_{\nu}f_{1\mu})^{2} 
\nonumber \\
 & & -\frac{1}{4}\left[\partial_{\mu}\vec{\rho}_{\nu}
                 -\partial_{\nu}\vec{\rho}_{\mu}
 + g (\vec{\rho}_{\mu} \times \vec{\rho}_{\nu}
      +\vec{a}_{1\mu}  \times \vec{a}_{1\nu})\right]^{2} 
\nonumber \\
 &  & -\frac{1}{4}\left[\partial_{\mu}\vec{a}_{1\nu}
                       -\partial_{\nu}\vec{a}_{1\mu}
 + g (\vec{\rho}_{\mu} \times \vec{a}_{1\nu}
      +\vec{a}_{1\mu}  \times \vec{\rho}_{\nu})\right]^{2}\,
\mbox{,}\label{vv}\\
\frac{m_1^2}{2}\, \mbox{Tr}[(L^{\mu})^{2}+(R^{\mu})^{2}] & = & 
\frac{m_1^2}{2}\, \left( \omega_\mu^2 + \vec{\rho}_\mu^{\,2} 
    + f_{1\mu}^2 + \vec{a}_{1\mu}^{\,2} \right) \, . \label{mv}
\end{eqnarray}
\end{subequations}
Note that the $\omega$-meson completely decouples 
from the dynamics. In order to have non-vanishing 
coupling of the $\omega$ to the other fields, we would 
have to include the Wess-Zumino-Witten term
\cite{Wess:1971yu,Kaymakcalan:1983qq}. Note that, had we 
included globally $U(N_f)_V \times U(N_f)_A$-invariant 
terms such as e.g.\ ${\rm Tr}\,|\Phi L^{\mu}+R^{\mu}\Phi|^2$, 
the $\omega$-meson would not decouple. 

Since the $\omega$-meson is protected by the 
$U(1)_V$ symmetry, it does not mix with the $f_1$. 
Consequently, we do not expect the $\omega$ to become 
degenerate in mass with the $f_1$ when chiral symmetry 
is restored. This is in contrast to the $\rho$ and the
$a_1$, which mix under $SU(2)_R\times SU(2)_L$ transformations: 
these mesons are expected to become degenerate in mass 
when chiral symmetry is restored.

In the phase where chiral symmetry is broken, the $\sigma$-field
assumes a non-vanishing expectation value, 
$\langle \sigma \rangle \equiv \phi = const. \neq 0$. 
In order to examine the fluctuations around the ground state,
we shift the $\sigma$-field by its vacuum expectation value, 
$\sigma \rightarrow \sigma + \phi$.
After this shift the covariant terms (\ref{vs}) read 
\begin{eqnarray}
\mbox{Tr}[(D^{\mu} \Phi )^{\dag}D_{\mu} \Phi ]&=&\;\;\;
\frac{1}{2} \left[\partial_{\mu}\sigma+g(\eta\, f_{1\mu} 
        + \vec{\pi}\cdot \vec{a}_{1\mu}) \right]^{2}\nonumber \\
 & & +\frac{1}{2}\left[\partial_{\mu}\eta
    -g(\sigma\, f_{1\mu}+\phi \, f_{1\mu}
     + \vec{a}_{0}\cdot \vec{a}_{1\mu})\right]^{2}
\nonumber \\ 
 & & +\frac{1}{2}\left[\partial_{\mu}\vec{a}_{0}
    +g(\vec{\rho}_{\mu}\times \vec{a}_{0}
     + \eta \, \vec{a}_{1\mu}+\vec{\pi}\, f_{1\mu})\right]^{2}
\nonumber \\
 & & +\frac{1}{2}\left[\partial_{\mu}\vec{\pi}
    -g(\vec{\pi} \times \vec{\rho}_{\mu}
     + \sigma \, \vec{a}_{1\mu}+\phi \, \vec{a}_{1\mu}
     + \vec{a}_{0}\, f_{1\mu})\right]^{2}\, .
\end{eqnarray} 
The shift of the $\sigma$-field leads 
(after an integration by parts) to the bilinear terms
$g \phi\, \eta \,\partial_\mu f_{1}^\mu$ and
$g \phi \, \vec{\pi} \cdot \partial_\mu \vec{a}_1^\mu$.
Physically, these terms correspond to a mixing
between the longitudinal component of the axial vector 
mesons and the Goldstone modes arising from 
chiral symmetry breaking.
If the mesons were true gauge fields, 
i.e., $m_1=0$, and if $c=H=0$, the theory would
be invariant under $U(N_f)_R \times U(N_f)_A$ {\em gauge\/}
transformations. In this case, the mixing could be removed
by an 't Hooft gauge-fixing term
\begin{equation}
{\cal L}_{GF}
  =  - \frac{1}{2\xi}(\partial_{\mu} \omega^{\mu})^{2}
     - \frac{1}{2\xi}(\partial_{\mu} \vec{\rho}^{\, \mu})^{2}
     - \frac{1}{2\xi}(\partial_{\mu}f_{1}^{\mu}
                 +\xi\, g\, \phi\, \eta)^{2}
     -\frac{1}{2\xi}(\partial_{\mu}\vec{a}_{1}^{\,\mu}+
      \xi\, g\,\phi\, \vec{\pi})^{2}\,\mbox{.}
\end{equation}
In unitary gauge, $\xi\rightarrow\infty$, the pseudoscalar 
particles (the Goldstone modes from spontaneously 
breaking $U(N_f)_V \times U(N_f)_A$ to $U(N_f)_V$) would 
become infinitely heavy, i.e, they are no longer 
dynamical degrees of freedom.  

Since $m_1 \neq 0$, the 't Hooft gauge-fixing procedure is
not at our disposal. We follow a method commonly used in the 
literature \cite{Gasiorowicz:1969kn,Ko:1994en,Pisarski:1994yp}, 
which is a redefinition of the axial fields,
\begin{subequations} \label{shift}
\begin{eqnarray}
f_1^\mu& = &{f'_1}^{\mu}
           + w\, \partial^{\mu}\eta\,, \\
\vec{a}_1^{\,\mu} & = & \vec{a}_1^{\, \prime \mu} 
           + w\, \partial^\mu \vec{\pi}\;,
\end{eqnarray}
\end{subequations}
where the new fields ared denoted by a prime (which is later
omitted) and $w$ is defined such that the 
bilinear terms mixing the axial vector and pseudoscalar 
fields are eliminated,
\begin{equation}
w = \frac{g \phi}{m^{2}_{1}+(g\phi)^{2}}\;.
\end{equation}
One could also perform a redefinition of the axial
vector fields using covariant derivatives
\cite{Gasiorowicz:1969kn,Ko:1994en},
\begin{subequations}
\begin{eqnarray}
f_1^\mu & = & {f'_1}^\mu 
         + w\, \left(\partial^{\mu} \eta 
          - g \,\vec{a}_1^{\,\mu} \cdot \vec{a}_0 \right)\;,\\
\vec{a}_1^{\, \mu} & = & \vec{a}_1^{\,\prime\mu} 
         + w\, \left[\partial^{\mu} \vec{\pi} 
              - g \left( \,\vec{\pi} \times  \vec{\rho}^{\, \mu}
              +g \, f_1^\mu \vec{a}_0\right) \right]\;.
\end{eqnarray}
\end{subequations}
Note, however, that the old fields appear in the 
covariant derivates and couple the set of equations. 
Solving for the old fields in terms of the new
one obtains
\begin{subequations} 
\begin{eqnarray}
f_1^\mu & = & \left( {f'_1}^\mu + w \left\{ \partial^\mu \eta
 - g  \vec{a}_0 \cdot \left[ \vec{a}_1^{\,\prime\mu} 
            + w \left( \partial^\mu \vec{\pi} 
    + g  \vec{\rho}^{\, \mu} \times \vec{\pi} \right) \right]
               \right\} \right)
           \left( 1 - g^2 w^2 \vec{a}_0^2 \right)^{-1}\;, \\
\vec{a}_1^{\, \mu} & = & \left( \vec{a}_1^{\,\prime\mu} 
                         + w \left\{\partial^\mu \vec{\pi}
   - g \left[\vec{\pi}  \times \vec{\rho}^{\, \mu}
               + \vec{a}_0 \left( {f'}_1^\mu 
               + w \, \partial^\mu \eta \right) \right]
             \right\} \right)
          \cdot 
         \left( 1 - g^2 w^2 \vec{a}_0 \vec{a}_0 \right)^{-1}\;.
\label{a_1old}
\end{eqnarray}
\end{subequations}
Note that the term $\vec{a}_0 \vec{a_0}$ 
on the right-hand side in Eq.\ (\ref{a_1old}) is a 
dyadic product, and consequently the corresponding term
in parentheses a matrix in isospin space, which has to
be multiplied from the left with the preceding isospin vector.
The ensuing set of equations couples different
isospin components of the old and new $a_1$ fields. 
All this can be avoided
by simply performing the redefinition with the partial
derivative, Eq.\ (\ref{shift}).
It should be noted that such complications did not arise
in previous treatments \cite{Gasiorowicz:1969kn,Ko:1994en},
because they did not consider the $\eta, \vec{a_0},$ 
and $f_1$ mesons. 
In that case, the redefinition of the $a_1$ field
simply reads $\vec{a}_1^{\, \mu} =  \vec{a}_1^{\,\prime\mu} 
         + w\, \left(\partial^{\mu} \vec{\pi} 
              + g \,\vec{\rho}^{\, \mu} \times \vec{\pi}
              \right) \;.$
It has been noted in Ref.\ \cite{Ko:1994en}
that this gives the correct seagull term when coupling the 
vector mesons to the photon. 
However, in this work we are 
not concerned with coupling the vector mesons to 
leptonic currents, and thus we restrict the consideration
to the simpler version (\ref{shift}) of the redefinition.
A different method to cope with the problem 
is to define a non-diagonal propagator \cite{Urban:2001ru}.

After the redefinition (\ref{shift}), the kinetic terms of the 
pseudoscalar mesons acquire a wave-function renormalization,
\begin{eqnarray}
\frac{1}{2} \left(\partial_{\mu}\pi_a\right)^2 \longrightarrow
\frac12 Z^{-2} \left(\partial_{\mu}\pi_a\right)^2\,\mbox{,}
\end{eqnarray}
where 
\begin{equation}
\label{z2}
Z^2 =\frac{m_1^2+\left(g\phi\right)^2}{m_1^2}\,.
\end{equation}
In order to have correctly normalized asymptotic states, 
we have to redefine the pseudoscalar fields,
\begin{equation} \label{redef}
\pi_a\longrightarrow Z\, \pi_a\,\mbox{.}
\end{equation}
After the redefinition of the 
pseudoscalar particles the Lagrangian reads 
\begin{eqnarray}
{\cal L}&=&\phantom{+}\frac12\left[\partial_{\mu}\sigma
                    +g Z \left(\eta f_{1\mu}
                    +w Z\eta\partial_{\mu}\eta 
                    +\vec{\pi}\cdot\vec{a}_{1\mu}
                    +w Z \vec{\pi}\cdot\partial_{\mu}
                         \vec{\pi}\right)\right]^2
\nonumber\\
        & & +\frac12\left[\left(\partial_{\mu}\eta\right)^2
            +g^2\left(\sigma f_{1\mu}
               + w Z \sigma\partial_{\mu}\eta
               +\vec{a}_0\cdot\vec{a}_{1\mu}
               +w Z \vec{a}_0\cdot \partial_{\mu}
                           \vec{\pi}\right)^2  \right.
\nonumber\\
 & & \phantom{+\frac 12(}\left. \frac{}{}
      -2g\left( Z \partial^{\mu}\eta-g\phi f_1^{\mu}
       - g\phi w Z \partial^{\mu}\eta\right)
        \left(\sigma f_{1\mu}+ w Z \sigma\partial_{\mu}
              \eta+\vec{a}_0\cdot\vec{a}_{1\mu}
            + w Z \vec{a}_0\cdot\partial_{\mu}\vec{\pi}
             \right)\right]
\nonumber\\
 & & + \frac 12\left[\partial_{\mu}\vec{a}_0
       +g\left(\vec{\rho}_{\mu}\times\vec{a}_0
       + Z \eta\vec{a}_{1\mu}
       + w Z^2\eta\partial_\mu\vec{\pi}
       + Z \vec{\pi}f_{1\mu} 
       + w Z^2 \vec{\pi}\partial_{\mu}\eta\right)\right]^2
\nonumber\\
 & & + \frac 12\left[\left(\partial_{\mu}\vec{\pi}\right)^2
       +g^2\left( Z \vec{\pi}\times\vec{\rho}_{\mu}
         +\sigma\vec{a}_{1\mu}
         + w Z \sigma\partial_{\mu}\vec{\pi}
         +\vec{a}_0f_{1\mu}
         + w Z \vec{a}_0\partial_{\mu}\eta\right)^2\right.
\nonumber\\
 & & \phantom{+\frac 12(}\left.\frac{}{}
      -2g\left( Z \partial^{\mu}\vec{\pi}
       -g\phi\vec{a}_{1}^{\mu}
       - g\phi w Z \partial^{\mu}\vec{\pi}\right)
     \cdot \left( Z \vec{\pi}\times\vec{\rho}_{\mu}
       +\sigma\vec{a}_{1\mu}
       + w Z \sigma\partial_{\mu}\vec{\pi}+\vec{a}_0f_{1\mu}
       + w Z \vec{a}_0\partial_{\mu}\eta
\right)\right]
\nonumber\\
 & & -\frac 14\left(\partial_{\mu}\omega_{\nu}
                   -\partial_{\nu}\omega_{\mu}\right)^2
     -\frac 14\left(\partial_{\mu}f_{1\nu}
                   -\partial_{\nu}f_{1\mu}\right)^2
\nonumber\\
& & - \frac 14\left\{\partial_{\mu}\vec{\rho}_{\nu}
                    -\partial_{\nu}\vec{\rho}_{\mu}
     +g\left[\vec{\rho}_{\mu}\times\vec{\rho}_{\nu}
            +\vec{a}_{1\mu}\times\vec{a}_{1\nu}
            +  w Z \left(\vec{a}_{1\mu} \times 
                        \partial_{\nu}\vec{\pi}
            +\partial_{\mu}\vec{\pi}\times\vec{a}_{1\nu}\right)
            +w^2 Z^2 \partial_{\mu}\vec{\pi}\times
              \partial_{\nu}\vec{\pi}\right]\right\}^2
\nonumber\\
& & -\frac 14\left\{\partial_{\mu}\vec{a}_{1\nu}
                   -\partial_{\nu}\vec{a}_{1\nu}
          +g\left[\vec{\rho}_{\mu}\times\vec{a}_{1\nu}
                 +\vec{a}_{1\mu}\times\vec{\rho}_{\nu}
                 + w Z \left(\vec{\rho}_{\mu}\times
                          \partial_{\nu}\vec{\pi}
                 +\partial_{\mu}\vec{\pi}\times
                   \vec{\rho}_{\nu}\right)\right]\right\}^2
\nonumber\\
& &  - \frac{1}{2} \left[m_0^2-c+3
 \left(\lambda_1+\frac{\lambda_2}{2}\right)\phi^2\right]\, 
\sigma^2 - \frac{Z^2}{2} \left[m_0^2+c+ 
 \left(\lambda_1+\frac{\lambda_2}{2}\right)\phi^2 \right]
\, \eta^2 \nonumber \\
&& - \frac{1}{2} \left[m_0^2+c+ 
   \left(\lambda_1+3\frac{\lambda_2}{2}\right)\phi^2 \right]
\, \vec{a}_0^2 
- \frac{Z^2}{2} \left[m_0^2-c+ 
   \left(\lambda_1+\frac{\lambda_2}{2}\right)\phi^2 \right]
\, \vec{\pi}^2
\nonumber\\
&&+\frac{m_1^2}{2}\left(\omega_{\mu}^2+\vec{\rho}_{\mu}^2\right)
+ \frac{m_1^2 + (g \phi)^2}{2} \left(f_{1\mu}^2
       +\vec{a}_{1\mu}^2\right)
\nonumber\\
&&-\frac{1}{4} \left(\lambda_1+\frac{\lambda_2}{2}\right)
\left(\sigma^2+ Z^2 \eta^2+\vec{a}_0^2
    + Z^2\vec{\pi}^2\right)^2
    -\left(\lambda_1+\frac{\lambda_2}{2}\right) \phi \sigma
 \left(\sigma^2+Z^2\eta^2+\vec{a}_0^2
      +Z^2\vec{\pi}^2\right)
\nonumber\\
&&-\frac{\lambda_2}{2}
\left[\left(\sigma\vec{a}_0+Z^2\eta\vec{\pi}\right)^2
+Z^2\vec{a}_0^2\vec{\pi}^2
-Z^2\left(\vec{a}_0\cdot\vec{\pi}\right)^2\right]
-\lambda_2 \phi \vec{a}_0
\cdot\left(\sigma\vec{a}_0+Z^2\eta\vec{\pi}\right) 
\nonumber \\
&& - V(\phi) \,\,\mbox{,}\label{lagr}
\end{eqnarray}
where
\begin{equation}
V(\phi) = \frac{1}{2}\, (m_0^2 -c)\, \phi^2 + \frac{1}{4}\,
\left( \lambda_1 + \frac{\lambda_2}{2} \right)\, \phi^4 
- h_0\phi 
\end{equation}
is the classical potential energy density.
In the derivation of Eq.\ (\ref{lagr}), we have exploited 
the fact that $\phi$ is the minimum of the potential 
energy density, i.e.,
\begin{equation} \label{mini}
0 = \frac{{\rm d} V}{{\rm d} \phi} = \left[m_0^2 - c + 
\left(\lambda_1 + \frac{\lambda_2}{2} \right) \phi^2 \right] 
\phi - h_0\;.
\end{equation}
From the Lagrangian (\ref{lagr}), one reads off the 
tree-level masses for the scalar, pseudoscalar, vector, 
and axial vector mesons,
\begin{subequations}
\begin{eqnarray}
m_\sigma^2 & = & m_0^2-c+
3\left(\lambda_1+\frac{\lambda_2}{2}\right)\phi^2\;, \\
m_\eta^2 & = & Z^2 
\left[ m_0^2+c+ \left(\lambda_1+\frac{\lambda_2}{2}\right)\phi^2
\right]\, , \\
m_{a_0}^2 & = & m_0^2+c+
\left(\lambda_1+3\frac{\lambda_2}{2}\right)\phi^2\, , \\
m_\pi^2 & = & Z^2\left[m_0^2-c+
\left(\lambda_1+\frac{\lambda_2}{2}\right)\phi^2
\right]\, , \\
m_\omega^2 & = & m_\rho^2 = m_1^2\,, \\
m_{f_1}^2 & = & m_{a_1}^2 = m_1^2 + (g \phi)^2\,.
\end{eqnarray}
\end{subequations}
Note the factor $Z^2$ in the definition of the (squared)
mass for the pseudoscalar particles.
We conclude that the wave-function renormalization
factor $Z$ in Eq.\ (\ref{z2}) is equal to the ratio of 
the tree-level masses of the $a_1$ and the $\rho$ meson, 
$Z \equiv m_{a_1}/m_\rho$. In the framework of this
model, the difference in mass of the two mesons is 
due to the Higgs effect.

Using the expression for the pion mass, we obtain from Eq.\
(\ref{mini}) that $h_0 = Z^{-2} m_\pi^2 \phi$. 
Using the renormalization of the pion field (\ref{redef}), 
the second Eq.\ (\ref{conserve}) reads
$\partial_\mu {\cal J}_{A}^{i\mu} = Z^{-1} m_\pi^2 \phi\, \pi_i$.
From the PCAC relation 
$\partial_\mu {\cal J}_{A}^{i\mu}= f_\pi m_\pi^2 \, \pi_i$ 
we then conclude that the vacuum expectation value of the scalar
field is
\begin{equation} \label{vev}
\phi = Z \,f_\pi \equiv f_\pi \, \frac{m_{a_1}}{m_\rho}\;.
\end{equation}
Note that the KSFR relation \cite{KSFR}
\begin{equation} \label{KSFR}
\frac{g^2 f_\pi^2}{m_\rho^2} = \frac{1}{2}
\end{equation}
makes a definite prediction for $Z$, and thus for the relation
between the mass of the $\rho$ and the $a_1$: 
Eqs.\ (\ref{z2}), (\ref{vev}), and (\ref{KSFR}) imply 
that $Z=\sqrt{2}$, or that
$m_{a_1} = \sqrt{2}\, m_\rho$. 
For $m_\rho \simeq 770$ MeV, this means that 
$m_{a_1} \simeq 1090$ MeV, somewhat smaller than the 
value quoted by the PDG \cite{Yao:2006px}, 
$m_{a_1} \simeq 1230$ MeV.

The vector and axial vector mesons are massive spin-one 
particles, i.e., they have three physical degrees of freedom. 
In order to have an invertible inverse tree-level propagator 
for these particles, we have to promote the unphysical 
fourth degree of freedom of the vector and axial vector
fields to a physical degree of freedom by adding a 
Stueckelberg term,
\begin{equation}
{\cal{L}}_{St}= - \frac{\xi}{2} \left[ 
     \left( \partial_{\mu}\omega^{\mu} \right)^2
    +\left(\partial_{\mu}f_{1}^\mu\right)^2
    +\left(\partial_{\mu}\vec{\rho}^{\,\mu}\right)^2
    +\left(\partial_{\mu}\vec{a}_{1}^{\, \mu}\right)^2\right]
\,\,\mbox{.}
\end{equation}
The unphysical degrees of freedom can be removed by 
taking the limit $\xi \rightarrow 0$ at the end of 
the calculation.

\section{Thermal Meson Masses}\label{IV}

In this section we investigate chiral symmetry restoration
at nonzero temperature in the framework of
the model (\ref{lagr}). We explicitly compute the condensate
and the meson masses as functions of temperature. 
As mentioned in the introduction, 
we use the 2PI resummation scheme, or 
$\Phi$-functional \cite{Baym:1962sx}, 
or CJT formalism \cite{Cornwall:1974vz}.
The generating functional for 2PI Green's functions
is
\begin{eqnarray}
\Gamma\left[\phi,G\right]=S\left[\phi\right]+
\half\tr\ln G^{-1}+\half\tr\left(D^{-1} G-1\right)
+\Gamma_2\left[\phi, G\right]\,\,\mbox{,}\label{gamma}
\end{eqnarray}
where $\phi$ and $G$ are the expectation values of the 
one- and two-point functions in the presence of 
external sources, $S\left[\phi\right]$ is the tree-level
action, $D^{-1}$ is the tree-level propagator and 
$\Gamma_2\left[\phi, G\right]$
is the sum of all 2PI vacuum diagrams with internal lines given
by $G$. If the system is translationally invariant, it
suffices to consider the effective potential 
$V=-T\Gamma/\Omega$, 
where $T$ is the temperature and $\Omega$ the $3$-volume of the
system. 

\begin{figure}
\begin{center}
\includegraphics[width=4cm]{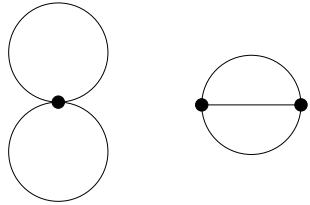}
\caption{Two-loop order diagrams in $V_2$.}
\label{plot_loop}
\end{center}
\end{figure}

\begin{figure}
\begin{center}
\includegraphics[width=4cm]{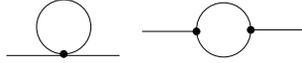}
\caption{1PI self-energy $\Pi$.}
\label{plot_self}
\end{center}
\end{figure}

For the Lagrangian (\ref{lagr}) the effective potential
is given by 
\begin{equation}
V\left[\phi,G_i\right]  = 
V(\phi)+I_{\sigma}+I_{\eta}+3I_{a_0}+3I_{\pi}+I_\omega + I_{f_1}
+3I_{\rho}+3I_{a_1}+V_2\left[\phi,G_i\right]
\,\,\mbox{,} \label{effpot}
\end{equation}
where $V_2 = - T \Gamma_2/ \Omega$, and
\begin{eqnarray}
I_i=\frac{1}{2}\int_k [\ln G_i^{-1}(k)
                      +{\cal{D}}_i^{-1}(k) G_i(k)-1]\;,
\end{eqnarray}
with the inverse tree-level propagators for scalar particles
${\cal D}_i^{-1} = -k^2 + m_i^2$, $i=\sigma,\eta,a_0,\pi$,
and for vector particles
${\cal D}_{i\; \mu \nu}^{-1} = -(k^2-m_i^2)g_{\mu\nu}+
(1-\xi)k_{\mu}k_{\nu}$, $i = \omega,f_1,\rho,a_1$.
The condensate $\phi$ and the full propagators $G_i$ 
are determined from the stationarity conditions for the 
effective potential (\ref{effpot}),
\begin{equation} \label{stat}
\frac{\delta V}{\delta \phi} = 0\;,\;\;\;\;
\frac{\delta V}{\delta G_i} = 0\;.
\end{equation}
The last equation is identical to the Dyson-Schwinger equation
for the full propagators,
\begin{equation} \label{schwinger}
G^{-1}_i = D^{-1}_i + \Pi_i\;,
\end{equation}
where the 1PI self-energy $\Pi_i$ is given by
\begin{equation} \label{self-energy}
\Pi_i =-2\frac{\delta V_2}{\delta G_i}\,\,\mbox{.}
\end{equation}
To lowest order in the loop expansion, i.e., to two-loop order,
two types of diagrams contribute to $V_2$. 
The diagram in Fig.\ \ref{plot_loop}, made of a single 
four-point vertex, is called the double-bubble diagram. 
The other one, made of two three-point vertices, 
is called the sunset diagram.
In Eq.\ (\ref{self-energy}), the functional derivative 
of $V_2$ with respect to a full propagator is 
diagrammatically equivalent to cutting a line
corresponding to that propagator. 
This gives rise to the self-energy diagrams
shown in Fig.\ \ref{plot_self}.
In contrast to the sunset diagram, the double-bubble 
diagram does not produce an imaginary part in
the 1PI self-energy when analytically continuing the 
latter to real energies. 
Thus, it gives only constant
(temperature-dependent) corrections 
to the mass and the wave-function renormalization constant of 
the particle. Neglecting the sunset terms is, 
of course, still a self-consistent approximation 
in the framework of the 2PI formalism. 
The advantage is that it tremendously 
reduces the numerical effort: instead of solving
a set of coupled complex integral equations for the self-energy
functions $\Pi(k_0, \vec{k})$ for a given temperature,
one only has to solve a set of coupled real
fixed-point equations for the masses and wave-function
renormalization constants. 

For the Lagrangian (\ref{lagr}), 
we explicitly list $V_2$ in double-bubble 
approximation in Eq.\ (\ref{V_2}) of App.\
\ref{app}. As detailed in that appendix, 
we make a further approximation to
simplify the structure of the vertices corresponding
to interactions of vector and axial vector mesons with
themselves and with pions. This approximation
yields degenerate dispersion relations for transverse and
longitudinal (axial-) vector degrees of freedom.
The resulting  
self-energies are given in Eqs.\ (\ref{selfenergy}), and the
set of coupled fixed-point equations in Eqs.\ (\ref{gapeq}).

\begin{figure}
\includegraphics[width=12cm]{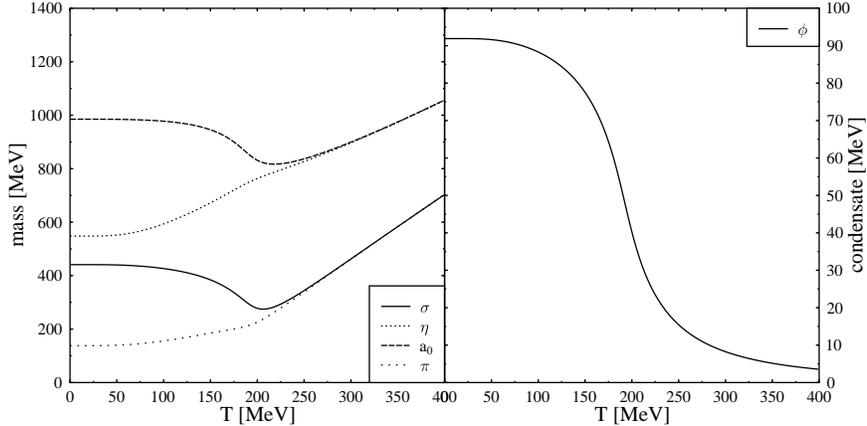}
\caption{Masses for scalar and pseudoscalar mesons (left panel)
and chiral condensate (right panel) as a function of temperature
for the case $g=0$, $m_\sigma=441$ MeV.}
\label{scal441}
\end{figure}

\begin{figure}
\includegraphics[width=12cm]{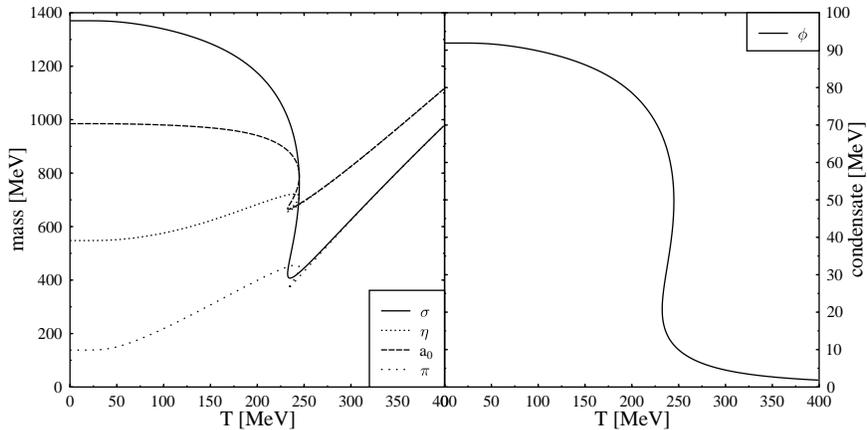}
\caption{The same as in Fig.\ \ref{scal441}, for 
$m_\sigma=1370$ MeV.}
\label{scal1370}
\end{figure}

The parameters of the model are fixed to the vacuum masses and 
the pion decay constant. Without vacuum fluctuations,	
vacuum and tree-level masses are identical, and we have
\begin{subequations}
\begin{eqnarray}
\lambda_1&=&\frac{1}{2(Z f_{\pi})^2} \left[  
m_\sigma^2-\left( \frac{m_\pi}{Z}\right)^2-m_{a_0}^2+
\left(\frac{m_\eta}{Z} \right)^2 \right]\; ,\\
\lambda_2&=&\frac{1}{(Z f_{\pi})^2}\left[ m_{a_0}^2-
\left(\frac{m_\eta}{Z} \right)^2\right]\; ,\\
m^2_0&=&\left(\frac{m_\pi}{Z}\right)^2
+\frac{1}{2} \left[ \left( \frac{m_\eta}{Z} \right)^2-m_\sigma^2
\right]\;, \\
c&=&\frac{m_\eta^2-m_\pi^2}{2Z^2}\;,\\
h_0&=&\frac{f_{\pi}m_\pi^2}{Z}\;,\\
m_1^2&=&m_\rho^2\;,\\
g^2&=&\frac{m_{a_1}^2-m_\rho^2}{(Zf_{\pi})^2}\;,
\end{eqnarray}
\end{subequations}
with
\begin{eqnarray}
m_\sigma&=&400-1400\,\mbox{MeV}\;,\nonumber\\
m_\eta&=&547.75\,\mbox{MeV}\;,\nonumber\\
m_{a_0}&=&985.1\,\mbox{MeV}\;,\nonumber\\
m_\pi&=&138.04\,\mbox{MeV}\;,\nonumber\\
m_\rho&=&768.5\,\mbox{MeV}\;,\nonumber\\
m_{a_1}&=&m_{f_1} = 1230\,\mbox{MeV}\;,\nonumber\\
f_{\pi}&=&91.9\,\mbox{MeV}\nonumber\,\, .
\end{eqnarray}
In Figs.\ \ref{scal441} and \ref{scal1370}
we show the temperature dependence of the 
condensate and the scalar meson masses without the influence of
vector mesons, i.e., setting $g=0$ (and $Z=1$).
In Fig.\ \ref{scal441}, we choose a vacuum sigma mass of 
$m_\sigma=441$ MeV \cite{Leutwyler:2006gz}. This leads to
a cross-over phase transition at $T_c\simeq 193$ MeV, in
good agreement with the lattice QCD calculations of Ref.\ 
\cite{Karsch:2007vw}.

Taking a vacuum sigma mass of $m_\sigma=1370$ MeV
(which is another candidate for a
$0^{++}$ state \cite{Yao:2006px}), 
the phase transition is of first order at a critical 
temperature of $T_c\simeq 238$ MeV, cf.\
Fig.\ \ref{scal1370}. Within this scenario with 
only scalar, but no vector mesons, we obtain a 
second-order phase transition for
a critical sigma mass of about $m_\sigma^* \simeq 965$ MeV 
at a temperature of $T_c \simeq 222$ MeV.

\begin{figure}
\includegraphics[width=12cm]{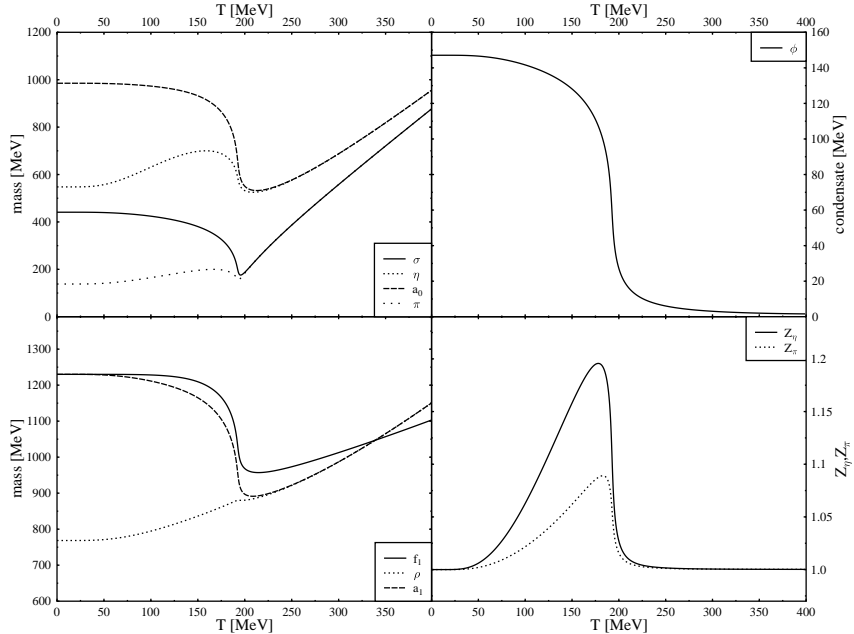}
\caption{Masses for scalar and pseudoscalar mesons 
(upper left panel), for vector and axial vector mesons 
(lower left panel), the chiral condensate (upper right panel), 
and the pion and $\eta$ wave-function renormalization constants 
as a function of temperature for the full model 
including vector and axial vector degrees of freedom; 
$m_\sigma=441$ MeV.}
\label{vec_scal441}
\end{figure}

\begin{figure}
\includegraphics[width=12cm]{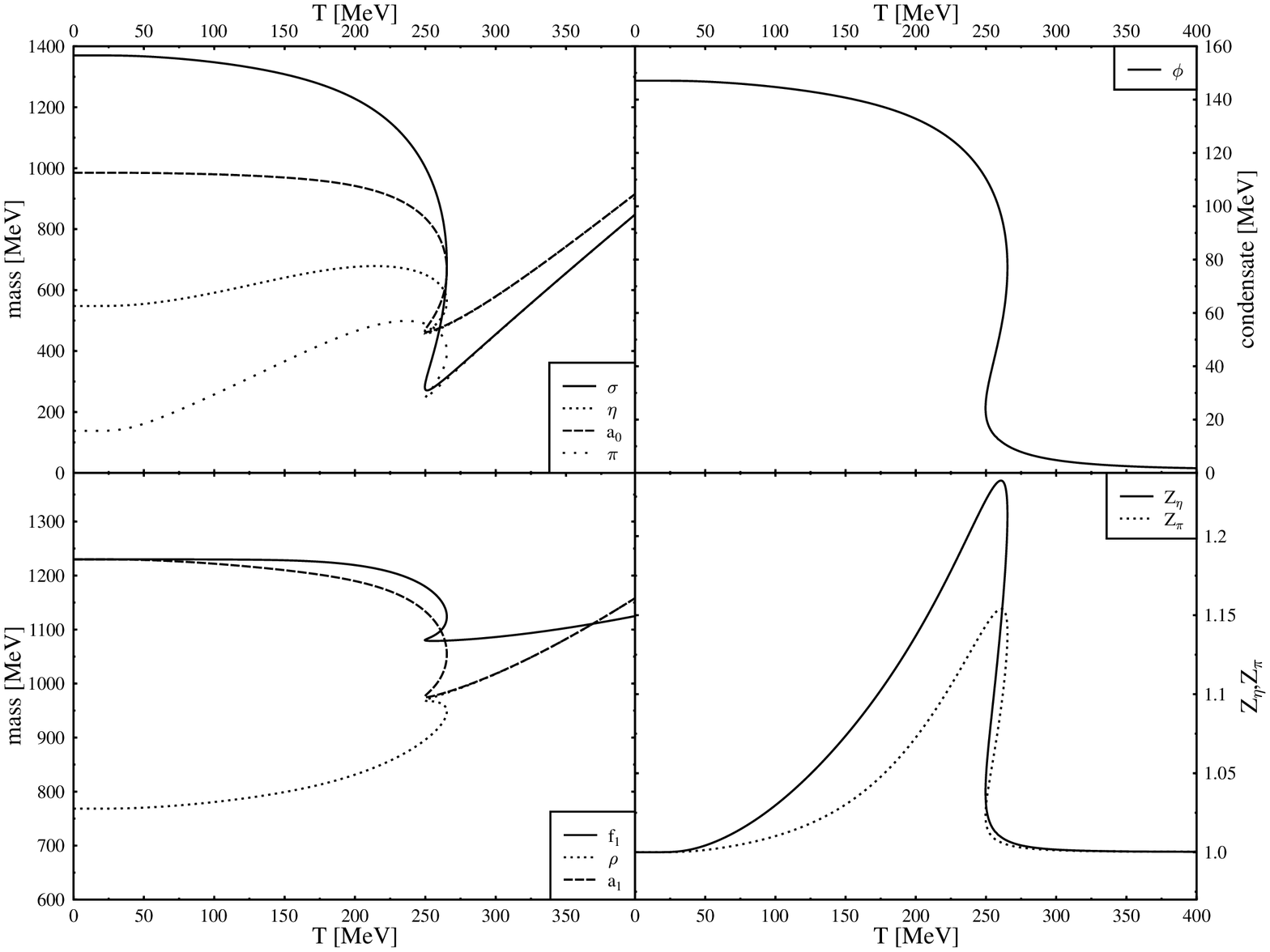}
\caption{Same as in Fig.\ \ref{vec_scal441}; 
$m_\sigma=1370$ MeV.}
\label{vec_scal1370}
\end{figure}

In Figs. \ref{vec_scal441} and \ref{vec_scal1370}, we
show the meson masses, the condensate, and the wave-function
renormalization constants as functions of temperature including
vector meson degrees of freedom.
For $m_\sigma=441$ MeV, cf.\ Fig.\ \ref{vec_scal441},
we observe a cross-over transition at about
the same temperature, $T_c \simeq 195$ MeV, 
as in the pure scalar case. However, the inclusion of vector 
mesons leads to a more rapid cross-over.
This is a typical effect within chiral linear 
$\sigma$ models when including more dynamical degrees of 
freedom, e.g., strange and charmed scalar mesons 
\cite{Roder:2003uz}.
Including vector mesons this effect is even stronger.

The increasing steepness of the cross-over transition 
indicates that the inclusion of vector mesons
brings one closer to a second-order critical point.
This can also be seen comparing the value of the $\sigma$ mass
at the respective transition temperature
in Figs.\ \ref{scal441} and \ref{vec_scal441}. In the latter, the
$\sigma$ meson is lighter by about 100 MeV. At a second-order
critical point, the $\sigma$ should become completely massless
\cite{Stephanov:1999zu}, indicating a diverging correlation 
length. 
Moving closer to a second-order critical point 
by the inclusion of vector degrees of freedom is advantageous 
from the following point of view. Let us consider
the second-order critical endpoint of the line
of first-order phase transitions associated with
the restoration of chiral symmetry, or shortly, the QCD 
critical point. Usually, in chiral 
models of QCD, this point is located at temperatures
which are too small and quark chemical potentials 
which are too large \cite{Stephanov:2007fk}
as compared to lattice QCD calculations \cite{Fodor:2004nz}.
Our calculations are performed at $\mu = 0$. The fact that
the inclusion of vector mesons brings one closer
to a second-order phase transition indicates that
the QCD critical point moves closer to the $T$ axis in
the $T - \mu$ plane.
We thus conjecture that the inclusion of vector mesons in linear
$\sigma$ models improves the agreement with lattice QCD.

Note that the $\rho$ mass increases, and the $a_1$ mass
decreases, as a function of temperature, 
until these mesons become
degenerate in mass beyond the chiral symmetry 
restoring transition. This is a necessary requirement 
for chiral symmetry restoration, as argued in 
Ref.\ \cite{Pisarski:1994yp}. The question remains, whether the 
mass of the $\rho$ increases to meet that of the $a_1$, 
or whether both decrease (or, for that matter, increase) 
with $T$, before becoming degenerate. In our model,
the increase of $M_\rho$ with $T$ is not surprising, 
because it consists of two separate contributions, 
the tree-level mass parameter $m_1$, 
and the contributions from thermal (tadpole) fluctuations.
While the former is constant as a function of temperature, 
one can show that the latter are always non-negative, 
therefore the mass of the $\rho$ has no choice but to 
increase with temperature.
This would be different, though, had we chosen to 
generate the tree-level mass dynamically through 
chiral symmetry breaking \cite{Ko:1994en}.
In this case, the tree-level contribution would decrease as
the condensate melts, while the thermal tadpole contributions 
would increase. (To our knowledge, this argument was first given 
in Ref.\ \cite{Brown:2005kb}.) 
As the condensate goes to zero in the process
of chiral symmetry restoration, only the
latter survive and lead to an increase of $M_\rho$ with $T$.
Nevertheless, in this scenario, at intermediate temperatures one
cannot exclude that the $\rho$ meson mass first decreases 
with $T$, assumes a minimum, and then increases again.
The minimum value of $M_\rho$ may well be zero, as advertised
by models with hidden gauge symmetry \cite{Harada:2005br}.

The pion and $\eta$ wave-function renormalization constants are
unity at $T=0$, and then increase by about 10 \% (for the pion)
or 20 \% (for the $\eta$) when approaching the 
chiral phase transition. They sharply drop back to one when the
transition temperature is exceeded.

In Fig.\ \ref{vec_scal1370} we show the thermal masses, the
condensate, and the wave-function renormalization for
$m_\sigma=1370$ MeV. In this case, there is a 
first-order phase transition at $T_c\simeq 258$ MeV. 
For the sake of
completeness we note that the second-order phase transition 
now occurs for a critical sigma mass 
$m_\sigma^* \simeq 795$ MeV at a temperature of 
$T_c \simeq 220$ MeV, i.e., at smaller values of 
$m_\sigma$ and $T_c$ than without vector mesons.


\section{Conclusions}\label{V}

In this work, we have investigated an effective theory for 
QCD with chiral $U(2)_R\times U(2)_L$ symmetry.
Besides scalar and pseudoscalar degrees of freedom, we included
vector and axial vector mesons.
These were coupled minimally to the scalar and 
pseudoscalar mesons.
The ensuing local $U(2)_R \times U(2)_L$ symmetry 
is explicitly broken to the corresponding global symmetry by
introducing a mass term for vector and axial vector mesons.
Without introducing a Wess-Zumino-Witten term, it turns out
that the $\omega$ meson completely decouples from the theory.

In the Goldstone phase, a non-vanishing chiral condensate
leads to terms bilinear in the axial vector and pseudoscalar
fields. Lacking the freedom to choose a convenient
(i.e., 't Hooft) gauge, 
in order to get rid of these terms one can perform
a redefinition of the axial vector fields with a subsequent 
redefinition of the pseudoscalar degrees of freedom. 

The resulting Lagrangian was employed to derive 
a set of coupled Dyson-Schwinger
equations via the 2PI formalism. 
In this work, we approximate the 2PI effective potential
by including only the so-called double-bubble diagrams.
This leads to constant, temperature-dependent 
shifts of the masses and, in the case of the pseudoscalar
mesons, the wave-function renormalization constants.

We did a control calculation without vector and
axial vector meson degrees of freedom in order
to study the influence of the latter on the chiral phase
transition. We found that, without vector mesons, and 
with a vacuum sigma mass $m_\sigma=441$ MeV, 
there is a smooth cross-over phase transition at a 
critical temperature of $T_c \simeq 193$ MeV. In the case
with vector mesons, this cross-over transition becomes
sharper, however, the transition temperature 
remains approximately the same, $T_c \simeq 195$ MeV. 

Increasing the mass of the sigma, the phase transition
becomes sharper and turns into a first-order transition
beyond a critical sigma mass $m_\sigma^*$. At
$m_\sigma^*$, the transition is second of order.
For the case without vector mesons, this happens
for $m_\sigma^* \simeq 965$ MeV, with a critical temperature
of $T_c \simeq 222$ MeV.
For the case with vector mesons, this happens
already for a smaller critical sigma mass,
$m_\sigma^* \simeq 795$ MeV with a critical temperature of 
$T_c \simeq 220$ MeV.
The fact that $m_\sigma^*$ becomes smaller when including
vector mesons can be understood from the fact that, for a
given $m_\sigma$, the transition was found to become
sharper as compared to the case without vector mesons.
This indicates that the inclusion of vector mesons
brings one closer to a second-order critical point.

This has consequences for the location of the QCD
critical point in the $T-\mu$ phase diagram. Without 
vector mesons, chiral models of QCD usually predict this
point at temperatures which are too small and at
quark chemical potentials which are too large as compared to 
lattice QCD calculations.
Since our above mentioned results are obtained at
$\mu = 0$, the inclusion of vector mesons would move 
this point closer to the $T$ axis and thus would improve
the agreement with the lattice results.

In our calculations, the mass of the $\rho$ meson is
a monotonously increasing function of temperature.
This is not surprising; in our model, the mass of
the $\rho$ essentially consists
of two contributions, the tree-level mass plus
thermal fluctuations from tadpole diagrams. The former
is always constant, while the latter are positive definite
and, in general, increasing functions of temperature.
A qualitatively different scenario would be one where
the tree-level mass is also generated dynamically
through spontaneous symmetry breaking.
A self-consistent treatment of this case within the 
Hartree approximation could be a topic of future studies.
However, the generic feature of chiral symmetry restoration,
namely that the $\rho$ and the $a_1$ become degenerate in mass
at the chiral phase transition, should still hold
for this case.

The ultimate goal is to find signatures for
chiral symmetry restoration in hot and dense nuclear
matter as created in heavy-ion collisions.
In this context, a promising observable is the invariant
mass spectrum of dileptons, since they carry direct
information from the hot and dense stages of the
evolution of the system. Of particular interest
is the contribution of the $\rho$ meson to the
dilepton spectrum, since it has a large (vacuum) decay
width and therefore decays inside the medium.
Changes of the spectral properties of the $\rho$ meson
potentially related to chiral symmetry restoration
can thus be directly seen in the dilepton invariant
mass spectrum.

With this in mind, the present study should 
be extended along several lines. 
In the double-bubble approximation
to the 2PI effective action, the quasiparticle
self-energy has no imaginary part.
Thus, in this approximation the quasiparticles cannot decay.
A natural next step is therefore
to include sunset-type diagrams in the 2PI effective action,
which lead to nonzero imaginary parts for the self-energy
of the quasiparticles and, in turn, to a nonzero 
decay width.
Another possible project is the self-consistent
calculation of meson masses for 
the gauged linear sigma model extended to
the three-flavor case \cite{Gasiorowicz:1969kn}.
From previous studies of dilepton production we
know that baryons play an important role
\cite{Rapp:1999ej}. Therefore, it is mandatory to include
nucleons into the present model.
Of course, a linear representation of chiral symmetry
requires to also add the chiral partner of the nucleon
as an explicit degree of freedom 
\cite{DeTar:1988kn,Wilms:2007uc}.
Since the chiral partner of the nucleon is most likely
heavier than the $\Delta$ resonance, the latter is
more abundant in a hot and dense system.
It is therefore unavoidable to also include
spin $3/2$-resonances and their chiral partners
\cite{Jido:1999hd}.

\section*{Acknowledgment}
The authors thank Hendrik van Hees, 
Robert Pisarski, Dirk R\"oder, J\"org Ruppert, 
J\"urgen Schaffner-Bielich, Igor Shovkovy,
and Detlev Zschiesche for valuable discussions.
S.S.\ thanks GSI Darmstadt for support through an
F{\&}E grant.

\appendix

\section{Effective potential and gap equations} \label{app}

In this appendix, we simplify our notation by denoting
full propagators of a given field by the field itself, 
for instance $G_\sigma(k) = \sigma(k)$, $G_\rho^{\mu \nu}(k) = 
\rho^{\mu \nu}(k)$ etc.
The trace over a vector propagator is denoted by
the corresponding field without Lorentz indices,
for instance $\rho^\mu_\mu (k) = \rho(k)$.
With this notation, the generating functional in
double-bubble approximation reads: 
\begin{eqnarray}
V_2&=&\frac14\left(\lambda_1+\frac{\lambda_2}{2}\right)
            \left\{3\left[\int_k\sigma(k)\right]^2
                  +3Z^4\left[\int_k\eta(k)\right]^2
                  +15\left[\int_k a_0(k)\right]^2
                  +15Z^4\left[\int_k\pi(k)\right]^2\right.
\nonumber\\
& &\phantom{\left(\frac{\lambda_1}{4}+\frac{\lambda_2}{8}\right)+}\left.
          +2Z^2\int_k\sigma(k)\int_l\eta(l)
          +6Z^2\int_k\sigma(k)\int_l\pi(l)
          +6Z^2\int_k\eta(k)\int_l a_0(l)\right\}
\nonumber\\
& &
    +\frac32\left(\lambda_1+\frac{3\lambda_2}{2}\right)
              \left[\int_k\sigma(k)\int_l a_0(l)
              +Z^4\int_k\eta(k)\int_l\pi(l)\right]
    +\frac92Z^2\left(\lambda_1+\frac{7\lambda_2}{6}\right)
                           \int_k a_0(k)\int_l\pi(l)
\nonumber\\
& &
    -\frac{g^2}{2}w^2Z^2\left(\int_kk^2\eta(k)
            \left\{\int_l\sigma(l)+Z^2\int_l\eta(l)
          +3\left[\int_l a_0(l)+Z^2\int_l\pi(l)\right]\right\}\right.
\nonumber\\
& &\phantom{-w^2Z^2\frac{g^2}{2}}\left.
           +3\int_kk^2\pi(k)\left[\int_l\sigma(l)+Z^2\int_l\eta(l)
           +\int_l a_0(l)+Z^2\int_l\pi(l)\right]\right)\nonumber\\
& &
          +\frac32g^2w^4Z^4\left\{\left[\int_kk^2\pi(k)\right]^2
              -\int_kk^\mu k^\nu\pi(k)\int_ll_\mu l_\nu\pi(l)\right\}
\nonumber\\
& &
+\frac{g^2}{2}\left\{\int_k f_1(k)
                   \left[\int_l\sigma(l)+Z^2\int_l\eta(l)
                       +3\int_l a_0(l)+3Z^2\int_l\pi(l)\right]
                   +6\int_k\rho(k)\left[\int_l a_0(l)
                                +Z^2\int_l\pi(l)\right]\right.
\nonumber\\
& &\phantom{\frac{g^2}{2}\,+}
                  +3\int_k a_1(k)\left[\int_l\sigma(l)+Z^2\int_l\eta(l)
                                  +\int_l a_0(l)+Z^2\int_l\pi(l)\right]
\nonumber\\
& &\phantom{\frac{g^2}{2}\,+}\left.
                  -6w^2Z^2\left[\int_k\rho^{\mu\nu}(k)
                  +\int_ka_1^{\mu\nu}(k)\right]
          \int_l \left(l^2 g_{\mu \nu} -l_\mu l_\nu\right)\pi(l)
                   \right\}
\nonumber\\
& &
+\frac{3}{2}g^2\left\{\left[\int_k\rho(k)\right]^2
                    -\int_k\rho^{\mu\nu}(k)\int_l\rho_{\mu\nu}(l)
                   +\left[\int_k a_1(k)\right]^2
                    -\int_k a_1^{\mu\nu}(k)\int_l a_{1\mu\nu}(l)\right.
\nonumber\\
& &\phantom{6\frac{g^2}{4}+}\left.
                    +2\left[\int_k\rho(k)\int_l a_1(l)
                    -\int_k\rho^{\mu\nu}(k)\int_l a_{1\mu\nu}(l)
\right]\right\}\,\,.
\label{V_2}
\end{eqnarray}

At this point we decompose the terms in $V_2$ derived 
from vectorial vertices in the following manner.  
Let $G_V^{\mu \nu}$ denote a vector or axial-vector propagator 
(with trace $G_V$, according to our above convention). We then
decompose a rank-2 tensor structure $B^{\mu \nu}$ as follows,
\begin{equation}
B^{\mu \nu} = \frac{1}{4}\, g^{\mu \nu}\, B + 
\delta B^{\mu \nu}\;,
\;\;\; \delta B^{\mu \nu} = B^{\mu \nu} - 
 \frac{1}{4}\, g^{\mu \nu} \, B\;,
\end{equation}
and apply this to $G_V^{\mu \nu}$ as well as
$k^\mu k^\nu$. We then neglect terms involving 
$\delta B^{\mu \nu}$. This results in the replacements
\begin{subequations}\label{approx}
\begin{eqnarray}
          \left[\int_k G_V(k)\right]^2
          -\int_k G_V^{\mu\nu}(k)\int_l G_{V\,\mu\nu}(l)
\,&\longrightarrow&\,\frac{3}{4}\left[\int_k G_V(k)\right]^2
\,\,,\\
          \int_k G_V(k)\int_l l^2 \pi(l)
          -\int_k G_V^{\mu\nu}(k)\int_l l_{\mu}l_{\nu} \pi(l)
\,&\longrightarrow&\,\frac{3}{4}\int_k G_V(k) \int_l l^2 \pi(l)
\,\,,\\
          \int_k k^2 \pi(k)\int_l l^2 \pi(l)-
\int_k k^{\mu}k^{\nu}
\pi(k)\int_l l_{\mu}l_{\nu}\pi(l)
\,&\longrightarrow&\,\frac{3}{4}\left[\int_k k^2 \pi(k)\right]^2
\,\,\mbox{.}
\end{eqnarray}
\end{subequations}
In this way, the vector and axial-vector meson self-energies
become proportional to $g^{\mu \nu}$, i.e., 
transverse and longitudinal self-energies become identical. 

The condensate equation with the approximated vectorial 
vertices reads
\begin{eqnarray}
h_0&=&\phi\left[\left(m_0^2-c\right)
   +\left(\lambda_1+\frac{\lambda_2}{2}\right)\phi^2
   +3\left[\left(\lambda_1+\frac{\lambda_2}{2}\right)
           \int_k\sigma\left(k\right)
   +\left(\lambda_1+3\frac{\lambda_2}{2}\right)
        \int_k a_0\left(k\right)\right]\right.\nonumber\\
  &&+\frac12\left\{\left[m_0^2+c
    +\left(\lambda_1+\frac{\lambda_2}{2}\right)\phi^2\right]u
    +2 Z^2 \left(\lambda_1+\frac{\lambda_2}{2}\right)
            \right\}\int_k\eta\left(k\right)
    \nonumber\\
  &&+\frac32\left\{\left[m_0^2-c
    +\left(\lambda_1+\frac{\lambda_2}{2}\right)\phi^2\right]u
    +2 Z^2 \left(\lambda_1+\frac{\lambda_2}{2}\right)
            \right\}\int_k\pi\left(k\right)\nonumber\\
  &&    +g^2\left[\int_k f_1\left(k\right)
    +3\int_k a_1\left(k\right)\right]
    +\frac32\int_k\pi\left(k\right)
       \left[v\left(\lambda_1+\frac{3\lambda_2}{2}\right)
                    \int_l\eta\left(l\right)
             +3u\left(\lambda_1+\frac{7\lambda_2}{6}\right)
                    \int_la_0\left(l\right)\right]
\nonumber\\
  &&+\frac14\left(\lambda_1+\frac{\lambda_2}{2}\right)
  \left(2u\left[\int_k\sigma\left(k\right)\int_l\eta\left(l\right)
                       +3\int_k\sigma\lek\int_l\pi\lel
                       +3\int_k\eta\lek\int_l a_0\lel\right]\right.
                                                    \nonumber\\
  &&\phantom{+\frac14\left(\lambda_1+\frac{\lambda_2}{2}\right)+}\left.
              +3v\left\{\left[\int_k\eta\left(k\right)\right]^2
                      +5\left[\int_k\pi\lek\right]^2\right\}\right)
                             \nonumber\\
  &&+\frac{g^2}{2}\int_k\eta\left(k\right)\left\{
     u_1\left[\int_ll^2\eta\left(l\right)
                +3\int_ll^2\pi\left(l\right)\right]
     +u\left[\int_lf_1\left(l\right)+3\int_la_1\left(l\right)\right]
     \right\}\nonumber\\
  &&+\frac{3}{2}g^2\int_k\pi\left(k\right)\left\{
       u_1\left[\int_ll^2\eta\left(l\right)
                +\int_ll^2\pi\left(l\right)\right]
      +u\left[\int_lf_1\left(l\right)+2\int_k\rho\left(l\right)
             +\int_l a_1\left(l\right)\right]\right\}\nonumber\\
  &&+\frac{g^2}{2}r\left\{\int_k\sigma\left(k\right)
        \left[\int_ll^2\eta\left(l\right)
               +3\int_ll^2\pi\left(l\right)\right]
         +3\int_ka_0\left(k\right)\left[
            \int_ll^2\eta\left(l\right)+\int_ll^2\pi\left(l\right)
            \right]\right\}\nonumber\\
  &&\left.
    +\frac94g^2r\int_kk^2\pi\left(k\right)\left[\int_l\rho\left(l\right)
                              +\int_la_1\left(l\right)\right]
     +\frac98g^2v_1\left[\int_kk^2\pi\left(k\right)\right]^2\right]
\,\,\mbox{,}\label{condensate}
\end{eqnarray}
where
\begin{subequations}
\begin{eqnarray}
u&=&\frac{1}{\phi}\frac{d (Z^2)}{d\phi}
  =\frac{2g^2}{m_1^2}\,\,,\\
v&=&\frac{1}{\phi}\frac{d (Z^4)}{d \phi}
  =\frac{4g^2\left(m_1^2+g^2\phi^2\right)}{m_1^4}\,\,,
\end{eqnarray}
\begin{eqnarray}
r&=&-\frac{1}{\phi}\frac{d (w^2 Z^2)}{d\phi}
  =-\frac{2g^2}{\left(m_1^2+g^2\phi^2\right)^2}\,\,,\\
u_1&=&-\frac{1}{\phi}\frac{d (w^2 Z^4)}{d\phi}
  =-\frac{2g^2}{m_1^4}\,\,,\\
v_1&=&\frac{1}{\phi}\frac{d (w^4 Z^4)}{d\phi}
  =\frac{4g^4\phi^2}{m_1^2\left(m_1^2+g^2\phi^2\right)^3}\,\,.
\end{eqnarray}
\end{subequations}

The approximation (\ref{approx}) has the distinct feature
that the vector-meson self-energies are proportional
to the metric tensor, $\Pi^{\mu \nu}_i = g^{\mu \nu} \, \Pi_i$,
$i=f_1, \rho, a_1$. 
The self-energies with the approximated vectorial vertices are
\begin{subequations}\label{selfenergy}
\begin{eqnarray}
\Pi_{\sigma}&=&\left(\lambda_1+\frac{\lambda_2}{2}\right)
                \left\{3\int_k \sigma(k) + Z^2\left[\int_k \eta(k)
                      + 3\int_k \pi(k)\right]\right\}
                +3\left(\lambda_1+3\frac{\lambda_2}{2}\right)
                    \int_k a_0(k)\nonumber\\
      & &-g^2\left\{w^2Z^2\left[\int_k k^2\eta(k)
                                +3\int_k k^2\pi(k)\right]
                -\int_k f_1(k)-3\int_k a_1(k)\right\}\,\,,\\
Z^{-2}\Pi_{\eta}(l)&=&Z^{-2}\Pi_{\eta}^{\star}
                      +Z^{-2}l^2\Pi_{\eta}^{\star\star}\nonumber\\
                   &=&\left(\lambda_1+\frac{\lambda_2}{2}\right)
                      \left[\int_k \sigma(k) + 3Z^2\int_k \eta(k)
                      + 3\int_k a_0(k)\right]
                       +3Z^2\left(\lambda_1+3\frac{\lambda_2}{2}\right)
                    \int_k \pi(k)\nonumber\\
 & &-g^2\left\{w^2Z^2\left[\int_k k^2\eta(k)
                                +3\int_k k^2\pi(k)\right]
                -\int_k f_1(k)-3\int_k a_1(k)\right\}\nonumber\\
 & & -g^2w^2l^2\left\{\int_k \sigma(k) + Z^2\int_k \eta(k)
                      +3\left[\int_k a_0(k)
                    +Z^2 \int_k \pi(k)\right]\right\}\,\,,\\
\Pi_{a_0}&=&\left(\lambda_1+\frac{\lambda_2}{2}\right)
                      \left[ Z^2\int_k \eta(k) + 5\int_k a_0(k)\right]
              +\left(\lambda_1+3\frac{\lambda_2}{2}\right)
                      \int_k \sigma(k)
                +  3Z^2\left(\lambda_1+7\frac{\lambda_2}{6}\right)
                      \int_k \pi(k)\nonumber\\
        & &-g^2\left\{w^2Z^2\left[\int_k k^2\eta(k)
                                  +\int_k k^2\pi(k)\right]
                -\int_k f_1(k)-2\int_k\rho(k)-\int_k a_1(k)\right\}\,\,,\\
Z^{-2}\Pi_{\pi}(l)&=&Z^{-2}\Pi_{\pi}^{\star}
              +Z^{-2}l^2\Pi_{\pi}^{\star\star}\nonumber\\
 &= & \left(\lambda_1+\frac{\lambda_2}{2}\right)
       \left[\int_k \sigma(k)+5 Z^2\int_k \pi(k)\right]
    + Z^2\left(\lambda_1+3\frac{\lambda_2}{2}\right)
               \int_k \eta(k)
    +3\left(\lambda_1+7\frac{\lambda_2}{6}\right)\int_k a_0(k)
  \nonumber\\
 & &-g^2\left\{w^2Z^2\left[\int_k k^2\eta(k)
                                  +\int_k k^2\pi(k)\right]
                -\int_k f_1(k)-2\int_k\rho(k)-\int_k a_1(k)\right\}\nonumber\\
 & & -g^2w^2l^2\left\{\int_k \sigma(k)+\int_k a_0(k)+
                               Z^2 \left[\int_k \eta(k) + \int_k \pi(k) 
                                 - \frac{3}{2}w^2\int_k k^2\pi(k)\right] 
            \right.\nonumber\\
 & & \phantom{-l^2g^2w^2+}\left.
                 +\frac{3}{2}\left[\int_k\rho(k)+\int_k a_1(k)\right]
                        \right\}\,\,,\\
\Pi_{f_1} & = & g^2\left\{\int_k\sigma(k)+3\int_k a_0(k)
                  + Z^2\left[ \int_k\eta(k)+
                   3\int_k\pi(k)\right]\right\} \,\,,\\
\Pi_{\rho}&=&2g^2 \left\{\int_k a_0(k)+ Z^2
                      \left[\int_k\pi(k)-\frac{3w^2}{4}
                       \int_k k^2\pi(k) \right]
                      +\frac{3}{4}\left[\int_k\rho(k) 
                      +\int_k a_1(k) \right]
                      \right\} \,\,,\\
\Pi_{a_1}&=&g^2 \left\{\int_k\sigma(k)
                     +\int_k a_0(k)+Z^2
                    \left[\int_k\eta(k)
                         +\int_k\pi(k)  
                         -\frac{3w^2}{2}\int_k k^2\pi(k)\right]
                 \right. \nonumber\\ 
                   && \hspace*{0.8cm}
                     \left. +\frac{3}{2}\left[\int_k\rho(k) 
                     +\int_k a_1(k)\right] \right\}
\,\mbox{.}
\end{eqnarray}
\end{subequations}
Finally, the coupled set of fixed-point equations for the
masses and wave-function renormalization constants is given by
\begin{subequations}\label{gapeq}
\begin{eqnarray}
        M_{\sigma}^2&=&m_\sigma^2+\Pi_{\sigma}\,\,,\\
          M_{\eta}^2&=&m_\eta^2 +\Pi_{\eta}^*\,\,,\\
         -Z_{\eta}^2&=&-1+\Pi_{\eta}^{**}\,\,,\\
           M_{a_0}^2&=&m_{a_0}^2+ \Pi_{a_0}\,\,,\\
           M_{\pi}^2&=&m_\pi^2  + \Pi_{\pi}^*\,\,,\\
          -Z_{\pi}^2&=&-1+ \Pi_{\pi}^{**}\,\,,\\
           M_{f_1}^2&=&m_{f_1}^2+\Pi_{f_1}\,\,,\\
          M_{\rho}^2&=&m_\rho^2+\Pi_{\rho}\,\,,\\
           M_{a_1}^2&=&m_{a_1}^2+\Pi_{a_1}\,\,,
\end{eqnarray}
\end{subequations}
which has to be solved together with Eq.\ (\ref{condensate}).
A consistent ansatz for the full propagators in momentum space
is
\begin{subequations}
\begin{eqnarray}
G_{\sigma,a_0}\left(k\right)&=&-\frac{1}{k^2-M_{\sigma,a_0}^2}
\,\,,\\
G_{\eta,\pi}\left(k\right)&=&-\frac{1}{Z_{\eta,\pi}^2}
\frac{1}{k^2-M_{\eta,\pi}^2/Z_{\eta,\pi}^2}\,\,,\\
G_{f_1,\rho,a_1}^{\mu\nu}\left(k\right)&=&
-\frac{1}{k^2-M_{f_1,\rho,a_1}^2}g^{\mu\nu}
-\frac{1}{M_{f_1,\rho,a_1}^2}
\left(\frac{1}{k^2-M_{f_1,\rho,a_1}^2/\xi}
-\frac{1}{k^2-M_{f_1,\rho,a_1}^2}\right)k^{\mu}k^{\nu}\mbox{.}
\end{eqnarray}
\end{subequations}
The tadpole integrals emerging after the Matsubara summation are
\begin{subequations}
\begin{eqnarray}
T_{\sigma,a_0}&=&\int_k-\frac{1}{k^2-M_{\sigma,a_0}^2}
=\int\frac{{\rm d}^3k}{(2\pi)^3}
\frac{1+2n_B(\omega_{\sigma,a_0})}{2\omega_{\sigma,a_0}}
\,\,,\\
T_{\eta,\pi}&=&\int_k-\frac{1}{Z^2_{\eta,\pi}}
\frac{1}{k^2-M_{\eta,\pi}^2/Z^2_{\eta,\pi}}
=\frac{1}{Z^2_{\eta,\pi}}
\int\frac{{\rm d}^3k}{(2\pi)^3}
\frac{1+2n_B(\omega_{\eta,\pi})}{2\omega_{\eta,\pi}}
\,\,,\\
T_{\eta,\pi}^*&=&\int_k-\frac{1}{Z^2_{\eta,\pi}}
\frac{k^2}{k^2-M_{\eta,\pi}^2/Z^2_{\eta,\pi}}
=\frac{M_{\eta,\pi}^2}{Z^4_{\eta,\pi}}
\int\frac{{\rm d}^3k}{(2\pi)^3}
\frac{1+2n_B(\omega_{\eta,\pi})}{2\omega_{\eta,\pi}}
\,\,,\\
T_{f_1,\rho,a_1}&=&\int_k\left[-\frac{4}{k^2-M_{f_1,\rho,a_1}^2}
-\frac{1}{M_{f_1,\rho,a_1}^2}
\left(\frac{1}{k^2-M_{f_1,\rho,a_1}^2/\xi}
-\frac{1}{k^2-M_{f_1,\rho,a_1}^2}\right)k^2\right]\nonumber\\
&=&3\int\frac{{\rm d}^3k}{(2\pi)^3}
\frac{1+2n_B(\omega_{f_1,\rho,a_1})}{2\omega_{f_1,\rho,a_1}}+
\frac{1}{\xi}\int\frac{{\rm d}^3k}{(2\pi)^3}
\frac{1+2n_B(\omega_{f_1,\rho,a_1}^\xi)}{
2\omega_{f_1,\rho,a_1}^\xi}\,\,,
\end{eqnarray}
\end{subequations}
where $\omega_i=(\vec{k}^2+M_i^2)^{1/2}$ 
for $i= \sigma, a_0,f_1, \rho, a_1$, while 
$\omega_i=(\vec{k}^2+M_i^2/Z_i^2)^{1/2}$ for
$i=\pi, \eta$, and
$\omega_i^\xi=(\vec{k}^2+M_i^2/\xi)^{1/2}$ for $i = \rho, a_1$;
$n_B(\omega)=[\exp(\omega/T)-1]^{-1}$ is the Bose-Einstein 
distribution. For the actual computation, we renormalize 
the tadpole integrals 
in such a way that divergent vacuum contributions
vanish. Other renormalization prescriptions (yielding nonzero
vacuum contributions) are possible, 
but do not lead to qualitatively different results  
\cite{Lenaghan:1999si}.
Note that, after taking the limit $\xi\rightarrow 0$, only
the three physical vector meson degrees of freedom
contribute to the tadpole integrals.

\end{document}